\documentclass[osajnl,preprint,showpacs,superscriptaddress,12pt]{revtex4-1}
\usepackage{epsfig}
\usepackage{amsmath}
\usepackage{amssymb}

\def\ao{Appl. Opt. }

\def\mnras{MNRAS\ }

\begin{document}

\title{Effects of dead time and after-pulses in photon detector on measured statistics of stochastic radiation}
\author{Victor Kornilov}\email{Corresponding author: victor@sai.msu.ru}
\affiliation{Moscow State University, Sternberg Astronomical Institute, Universitetsky pr-t, 13, Moscow, Russia}

\begin{abstract}

Many physical experiments require analysis of the statistics of fluctuating radiation. In the case of an ideal single-photon detector, the contribution of photon noise to the statistics of the registered signal has been thoroughly examined. However, practical photon counters have a dead time, leading to miscounting of certain true events, and sometimes the counters generate false after-pulses.

This study investigate the impact of these two effects, and it presents the theoretical relations between the statistical moments of the radiation and the registered counts while also accounting for dead time and the probability of after-pulses. Expressions for statistical moments of any order are obtained on the basis of the generalized Poisson distribution (GPD). For counters with paralyzable dead time, alternative relations for the mean and variance are derived using generally accepted formulas.

As an example, the measurements of stellar scintillation and the result of simple experiment are considered. The results of the experimental verification of the theoretical expression confirm the need to account for the non-ideal nature of detectors in almost all similar measurements.
\end{abstract}

%\ocis{(030.5260) Photon counting; (030.5290) Photon statistics; (290.5930) Scintillation; (000.5490) Probability theory, stochastic processes, and statistics}

\maketitle

\section{Introduction}

When measuring the intensity of optical radiation, the registered signal is inevitably burdened with photon (Poisson) noise, and the statistics concerning the received signal differ from those of incident radiation. This noise is not additive and in the case of fluctuating incident radiation its power changes depending on the radiation's instantaneous intensity. The problem concerning the contribution of photon noise in the measured statistical characteristics is considered a classical problem that has been solved many years ago.

Particular interest in the problem was triggered during the interpretation of the non-classical interference of light, study of Raman scattering, and photon correlation spectroscopy (see for e.g., \cite{Mandel1958,Jakeman1980TCP,Kuzmany2009}). Calculating the contribution of photon noise is important to several experimental works, particularly those that involve the photon counting method in low-light applications.

It has been shown that the registered mean, variance, and correlations are related with the corresponding set of statistics for stochastic radiation in a simple manner. A simple relationship called the Mandel relation \cite{Mandel1959,Parry1980,Cummins1974} connects the moments of any order of fluctuating intensity with the corresponding factorial moments of photon counts.

The Mandel relation is a convenient expression that has been repeatedly used, including in the study of stellar scintillation statistics \cite{Parry1980,Jakeman1978,Dainty1982,Stecklum1985}. However, this relation is valid for an  ideal detector while practical photoelectric detectors are not ideal.

Firstly, the photon-counting method involves nonlinearity caused by the presence of dead time in the registration chain. For example, in the measurements of stellar scintillation with the Multi-aperture scintillation sensor (MASS) \cite{2003SPIE}, accounting for nonlinearity is an essential step in signal processing \cite{2003MNRAS}, because in order to achieve the necessary statistical accuracy, we are forced to choose bright stars.

The attenuation of the light flux is not a solution, since in this case, a well-known effect that a single-photon detector may have after-pulses of different nature, becomes more noticeable. This effect also distorts the statistics of registered events. The after-pulse effect has not thus far been properly studied from the theoretical viewpoint, and ill-grounded empirical arguments are used to describe the effect.

In this study, we present analogues of the Mandel relation, which connect the moments of a stochastically fluctuating light flux (hereafter reffered to as scintillation) with the factorial moments of the registered counts for the case of noticeable nonlinearity of any given photon counter. The expressions are derived for both non-paralyzable and paralyzable dead time, i.e., counters of both the first (I) and the second (II) types. The relations consider the effect of after-pulses as well.

Experimental results confirming our theoretical conclusions are presented in Sec.~\ref{sec:exper}. The applicability of the relations is tested using actual  measurements of stellar scintillation. Finally, we present simple algorithms to solve the practical problem of the determination of the statistical moments of scintillation from measured data.

\section{Theory}

\subsection{Photocount statistics for ideal detector}

The measurement of statistical characteristics involves a large number of exposures with a exposure time $\Delta t$ over a time interval $T$. Thus, the number of discrete counts is given as $N = T/\Delta t$. Each count $x$ is a random variable, possibly correlated with adjacent counts. We assume that the scintillation is a stationary random process, and that it is described by the probability distribution $f(S; s^2)$ of the scintillation $S = I/\langle I \rangle$ which is the normalized flux incident on the detector. The value of $S$ is essentially non-negative, its expectation $\mathsf{E}[S] = 1$, and its variance $\mathsf{Var}[S] = s^2$ is the scintillation index.

In the ideal case, the probability to detect $x$ output events for an input flux $\xi S$ is described by the Poisson distribution $P(x; \xi S)$ (PD). The total probability of the registration of $x$ events with a mean flux $\xi$ can be evaluated using the expression
\begin{equation}
\phi(x;\xi, s^2)=\int\limits_{0}^{\infty}P(x;\xi S)f(S; s^2)\mathrm{\,d}S.
\label{eq:probab1}
\end{equation}
This description in the form of a doubly stochastic process (Cox process) is commonly accepted. Alternatively, the inhomogeneous Poisson process is used when the count rate is a smoothly varying function of time (see for e.g., \cite{Vannucci1981}). This approach is more convenient for deterministic light signals, but of little use in scintillation studies.

Let us calculate the mean value of $x$, which is distributed in accordance with the function $\phi(x; \xi, s^2)$, directly from its definition.
\begin{equation}
\mathsf{E}[x] = \sum\limits_{x=0}^{\infty} x\, \phi(x;\xi,s^2) = \int\limits_{0}^{\infty}\sum\limits_{x=0}^{\infty} x\,P(x,\xi S)\,f(S; s^2)\mathrm{\,d}S = \int\limits_{0}^{\infty}\xi S\,f(S; s^2)\mathrm{\,d}S = \xi.
\label{eq:mu1}
\end{equation}
This expression confirms the interpretation of the parameter $\xi $ as the mean value of the input flux of photon events.

To calculate the variance $\mathsf{Var}[x]$, we use the fact that for the PD, the factorial moment $\mu_\mathrm{[r]} = \theta^r = \mu_\mathrm{[1]}^r$, where $r$ denotes the order of the moment. The factorial moment $\alpha_\mathrm{[r]}$ of the registered count $x$ can be written as
\begin{equation}
\alpha_\mathrm{[r]} = \sum\limits_{x=0}^{\infty} (x)_r \int\limits_{0}^{\infty} P(x;\xi S)\,f(S; s^2)\mathrm{\,d}S,
\label{eq:fmom_calc0}
\end{equation}
where $(x)_r$ denotes the Pochhammer symbol for the lower factorial. After summation under the integral and replacing the sum with the expression for $\mu_\mathrm{[r]}$, we obtain
\begin{equation}
\alpha_\mathrm{[r]} = \int\limits_{0}^{\infty} (\xi S)^r f(S; s^2)\mathrm{\,d}S %\\= \xi^r \int\limits_{0}^{\infty} S^r f(S; s^2)\mathrm{\,d}S 
= \xi^r\,\nu^{\prime}_r,
\label{eq:mandell}
\end{equation}
where $\nu^{\prime}_r$ denotes the raw moment of order $r$ of the scintillation $S$. This equation is called the Mandel relation. Applying the trivial formulas $\alpha_\mathrm{[2]} = \mathsf{E}[x^2] - \mathsf{E}[x]$, $\mathsf{Var}[x] = \mathsf{E}[x^2] - (\mathsf{E}[x])^2$, and $s^2 = \nu^{\prime}_2 - 1$, we obtain the following well-known expression:
\begin{equation}
\mathsf{Var}[x] = \xi + \xi^2 s^2.
\label{eq:mu2}
\end{equation}

Equations (\ref{eq:mu1}) and (\ref{eq:mu2}) explain why sometimes instead of the process with the distribution given by Eq.~(\ref{eq:probab1}), Poisson noise is considered incorrectly as additive noise with power $\xi$ superimposed on the signal with mean $\xi$ and variance $\xi^2s^2$. In this interpretation, the scintillation has a very different distribution, although has the same the first two moments. Higher moments, which can be calculated by Eq.~(\ref{eq:mandell}), demonstrate the difference between these interpretations.

Equation~(\ref{eq:mandell}) can also be obtained from the representation of the probability generating function (PGF) $\gamma_x(t)$ through the moment generating function (MGF) $M_S(u)$ for the $f(S; s^2)$ distribution as below:
\begin{equation}
\gamma_x(t) = \sum\limits_{x=0}^{\infty} x^t \int\limits_{0}^{\infty} P(x;\xi S)\,f(S; s^2)\mathrm{\,d}S = \int\limits_{0}^{\infty} e^{S\xi(t-1)} f(S; s^2)\mathrm{\,d}S = M_S(\xi t - \xi),
\label{eq:fmgf}
\end{equation}
where $e^{\xi S\,(t-1)} = \sum x^t P(x; \xi S)$ over all values of $x$ is the Poisson PGF with the parameter $\xi S$. Since $x$ is a non-negative integer, $\gamma(t)$ is also the generating function of factorial moments. The relation between moments can be obtained in the usual way.

\subsection{Generalized Poisson distribution}

In the case of nonlinear detectors, the distribution $P(x; \xi S)$ is no longer a Poisson distribution. As we have shown in \cite{2008ARep}, an adequate description of registered events with a counter having a dead time $\tau_1$ is achieved by using the generalized Poisson distribution (GPD), introduced in \cite{Consul1973}. This distribution is the two-parameter distribution and it is usually written as
\begin{align}
P(k;\theta,\lambda) &=
\frac{\theta(\theta+k\lambda)^{k-1}e^{-\theta-k\lambda}}{k!}, \quad k = 0\dots n \notag \\
P(k;\theta,\lambda) &=0, \qquad k > n
\label{eq:gpd1}
\end{align}
where $\theta > 0$, and the parameter $\lambda$ defines the truncation of $n$ from the inequality $\max(-1, -\theta/n) \le \lambda < 1$. When $\lambda = 0$, the distribution reduces to an ordinary PD with the parameter $\theta$. The expectation (mean) and variance of GPD are respectively equal to
\begin{equation}
\mathsf{E}[k] = \frac{\theta}{1-\lambda}, \quad \mathsf{Var}[k] = \frac{\theta}{(1-\lambda)^3}.
\label{eq:gpd_mom}
\end{equation}

In our case, the parameter $\theta$ denotes the flux of input events $\xi S$, and the parameter $\lambda < 0$. For counter type I, $\lambda = -\tau\xi S$, and for the counter with paralyzable dead time, $\lambda = 1 - e^{\tau\xi S}$ \cite{2008ARep}. The parameter $\tau$ is the nonlinearity parameter, or dead time, reduced to the interval $\Delta t$, i.e., $\tau = \tau_1/\Delta t$. In the situations discussed below, $\tau_1 \ll \Delta t$ and hence $\tau \ll 1$.

From Eq.~\ref{eq:gpd_mom}, it follows that modulo $\lambda$ is the ratio of the number of missing events to the number of registered ones. The truncation has a formal character as $n$ is always considerably more than $\mathsf{E}[k]$. The parameter $\lambda$ can be simply obtained from the non-Poisson parameter $p = \mathsf{Var}[x]/\mathsf{E}[x] = (1 - \lambda)^{-2}$, which is usually measured during detector tests \cite{2003MNRAS}.

The generating functions for the GPD are known, and thus, the PGF can be written \cite{Ambagaspitiya1994} as:
\begin{equation}
\gamma(t) = e^{\theta(-W(-\lambda e^{-\lambda}t)/\lambda - 1)},
\label{eq:fgf}
\end{equation}
where $W(x)$ denotes the Lambert function (see for e.g., \cite{Corless1997}). It is easy to verify that $\gamma(1) = 1$, and in the limit $\lambda \to 0$, it coincides with the Poisson PGF.

The method applied in the case of Eq.~(\ref{eq:fmgf}), cannot be repeated directly, since in the discussed case, $\lambda$ is dependent on $\theta$, that does not allow to present Eq.~(\ref{eq:fgf}) as the scintillation MGF with compound argument. A relation similar to Eq.~(\ref{eq:fmgf}) can only be obtained by neglecting the variation of $\lambda$ with light intensity. The setting a certain value of $\lambda$ leads to a linear approximation, corresponding to the local slope of the nonlinear curve at this chosen value. However, it is known that in the case of strong scintillation, the maximal intensity exceeds the average flux by an order of magnitude. Therefore, it is required to perform a detailed analysis taking into account the curvature of the nonlinearity function, is desired.

In the following sections, the nonlinearity for constant flux is refered as the {\it static nonlinearity}. The effect, that caused by the fluctuations of light intensity and curvature of nonlinearity function, is reffered to as the {\it dynamic nonlinearity}.

\subsection{Factorial moments of GPD}
\label{sec:facmomGPD}

Using Eq.~(\ref{eq:fgf}), we can calculate any factorial moment $\mu_\mathrm{[r]}$ for the GPD. In the normalized form, the first four moments can be written as follows:
\begin{align}
\mu_\mathrm{[1]} &= \frac{\theta}{1-\lambda}, \notag \\
\frac{\mu_\mathrm{[2]}}{\mu_\mathrm{[1]}^2} &= 1 +\frac{\lambda\,(2-\lambda)}{\theta\,(1-\lambda)}, \label{eq:fmomGPD} \\
\frac{\mu_\mathrm{[3]}}{\mu_\mathrm{[1]}^3} &= 1 + \frac{3\,\lambda\,(2-\lambda)}{\theta\,(1-\lambda)} + \frac{\lambda^2(2\,\lambda^2-8\,\lambda+9)}{\theta^2(1-\lambda)^2}, \notag \\
\frac{\mu_\mathrm{[4]}}{\mu_\mathrm{[1]}^4} &= 1 + \frac{6\,\lambda(2-\lambda)}{\theta\,(1-\lambda)} +\frac{\lambda^2(11\,\lambda^2-44\,\lambda+48)}{\theta^2\,(1-\lambda)^2} + \dots \notag
\end{align}
Since $\lambda/\theta \approx -\tau$, the second term on the right-hand side is of the order $\tau$, the third is of the order of $\tau^2$, etc. In general, the numerical coefficient of the second term is equal to $r\,(r-1)/2$ for a moment of the order $r$. For a typical dead time of $\tau_1 = 10$--30~ns and exposure longer than 3~$\mu$s ($\tau \le 0.01$), we can ignore the third and subsequent terms. In most situations, only the first term is needed, but for higher moments, the factor $r\,(r-1)$ may be $\gtrsim$100 and the second term can be significant despite the small value of $\tau$.

We restrict our analysis to the first two terms, and represent the second term in the form $\omega r\,(r-1)\,\tau$, where the coefficient $\omega$ depends on the dead time type; in either case, it differs from unity by a factor less than $\tau\xi$. Subsequently,  $\mu_\mathrm{[r]}$ can be written as
\begin{equation}
\mu_\mathrm{[r]} = \mu_\mathrm{[1]}^r \left(1 - \omega r\,(r-1)\,\tau\right) = G_r\,\mu_\mathrm{[1]}^r.
\label{eq:fmom_appr}
\end{equation}
Since $\omega$ is multiplied by the small parameter $\tau$, it may be set to 1. For brevity, we denote the factor $1 - r\,(r-1)\,\tau$ as $G_r$.

\subsection{Effect of after-pulses on statistics}
\label{sec:afterpulse}

The relations in the previous section are valid for the case where all registered events are independent. However, in actual detectors, after-pulses of various natures can be observed, when one extraneous pulse appears with a probability $q \ll 1$, subsequent to the pulse corresponding to a photon event.

It is known that binomial selection applied to a random Poisson value $x$ leads also to the PD of variable $x_1$ with the parameter $q\theta$. The unselected part $x_2$ is also Poissonian with the parameter $(1-q)\,\theta$, and it is statistically independent from $x_1$. Let us assume that $x_1$ represents the events that give rise to after-pulses, and hence, instead of $x_1$, we will register $2x_1$. Obviously, in this case $\mathsf{E}[x] = \theta + q\theta$. The variance $\mathsf{Var}[x_2] = (1-q)\theta$, and for the sample $x_1$ with after-pulses, $\mathsf{Var}[x_1] = 4\,q\theta$. The total variance of $x$ amounts to $\mathsf{Var}[x] = (1-q)\theta + 4\,q\theta = \theta + 3\,q\theta$. As can be seen, an excess of variance appears, i.e. the process is not quite Poissonian, and the non-Poissonian factor $p = \mathsf{Var}[x]/\mathsf{E}[x] \approx 1 + 2\,q$.

In terms of the PGF, this Poisson process with after-pulses is written very simply as
\begin{equation}
\gamma_x(t) = \gamma_{x_2}(t)\gamma_{x_1}(t^2) = e^{(1-q)\theta(t-1)+q\theta(t^2-1)}.
\label{eq:ideal-after}
\end{equation}
Using this expession, we can obtain the factorial moments of the process as
\begin{align}
\mu_\mathrm{[1]} &= \theta(1+q), \\ 
\mu_\mathrm{[r]} &= \theta^r(1+q)^r + \theta^{r-1}r\,(r-1)\,q\,(1+q)^{r-2} + O(q^2), \notag
\label{eq:after_moms}
\end{align}
or in the normalized form as
\begin{equation}
\frac{\mu_\mathrm{[r]}}{\mu_\mathrm{[1]}^r} = 1 + \frac{r(r-1)q}{\mu_\mathrm{[1]}(1+q)} + O(q^2/\mu_\mathrm{[1]}^2).
\label{eq:reatr}
\end{equation}
From this formula, it follows that if the light intensity is decreasing, the normalized factorial moments increase infinitely, as opposed to the purely Poissonian case. When $\mu_\mathrm{[1]} < q$, the terms of $O(q^ 2/\mu_\mathrm{[1]}^2)$ also begin to influence the moments with $r > 3$.

The distortion of statistics caused by after-pulses leads to the invalidity of the Mandel relation (\ref{eq:mandell}). In its right-hand side, the factor $(1 + q)^r$ appears before the moment $\nu^{\prime}_r$, and the term $r(r-1)\,q\,(1 + q)^{r-2}\nu^{\prime}_{r-1}$ is another additional term.

The combined effect of detector nonlinearity and after-pulses can be estimated by constructing the appropriate PGF:
\begin{equation}
\gamma_x(t) = e^{(1-q)\theta(-W(-\lambda e^{-\lambda}t)/\lambda - 1)+q\theta(-W(-\lambda e^{-\lambda}t^2)/\lambda - 1)}.
\label{eq:GPD-after}
\end{equation}
Using this function, we calculated the first few factorial moments:
\begin{align}
\mu_\mathrm{[1]} &= \frac{\theta(1+q)}{1-\lambda}, \notag \\
\frac{\mu_\mathrm{[2]}}{\mu_\mathrm{[1]}^2} &= 1 \!+\! \frac{\lambda(2-\lambda)}{\mu_\mathrm{[1]}(1-\lambda)^2}\!+\!\frac{2q}{\mu_\mathrm{[1]}(1-\lambda)^2(1+q)},\label{eq:jointq} \\
\frac{\mu_\mathrm{[3]}}{\mu_\mathrm{[1]}^3} &= 1 \!+\! \frac{3\lambda(2-\lambda)}{\mu_\mathrm{[1]}(1-\lambda)^2} \!+\!\frac{6q}{\mu_\mathrm{[1]}(1-\lambda)^2(1+q)} \!+\! O\Bigl(\frac{\lambda^2}{\theta^2}\Bigr).
\notag
\end{align}
Evaluating the following moments, we ensured that the moment of order $r$ can be presented in the form
\begin{equation}
\frac{\mu_\mathrm{[r]}}{\mu_\mathrm{[1]}^r} = 1 + \frac{r(r-1)\lambda(2-\lambda)}{2\,\theta(1-\lambda)(1+q)} +\frac{r(r-1)q}{\theta(1-\lambda)(1+q)^2} + O\Bigl(\frac{\lambda^2}{\theta^2}\Bigr)+O\Bigl(\frac{q^2}{\theta^2}\Bigr).
\label{eq:momGPD-after}
\end{equation}

This expression satisfies two extreme cases: the absence of dead time (perfect linearity) and complete absence of after-pulses. As in Sec.~\ref{sec:facmomGPD}, we replace the first two terms with the coefficient $G_r \approx 1$ while neglecting the emerged factor $1 + q$. The third term at high fluxes introduces a small contribution, and therefore we can ignore distinction of $(1 - \lambda)^2$ from unity after multiplying Eq.~(\ref{eq:momGPD-after}) by $\mu_\mathrm{[1]}^r$. Denoting $H_r = r\,(r-1)\,q/(1 + q)$, we finally obtain
\begin{equation}
\mu_\mathrm{[r]} = G_r\,\mu_\mathrm{[1]}^r + H_r\,\mu_\mathrm{[1]}^{r-1}.
\label{eq:fmom_appraft}
\end{equation}
The coefficient $H_r$ for $r = 1$ is equal to zero; the increasing of the order leads to an almost quadratic growth of the coefficient. We recall that these are the factorial moments of the distribution $P(x; \theta)$.

\subsection{Factorial moments of registered counts}
\label{sec:registered}

To proceed further to the statistical moments of registered counts in the case of scintillation, at first we expand $\mu_\mathrm {[r]}$ (\ref{eq:momGPD-after}) in series powers of $\tau$. For counter type I, $\mu_\mathrm{[1]} = (1 + q)\xi S/(1 + \tau\xi S)$ and the expansion is given as
\begin{equation}
\mu_\mathrm{[r]} = G_r(1+q)^r\sum_{k=0}\frac{r^{(k)}}{k!} (\xi S)^{r+k}(-\tau)^k + H_r(1+q)^{r-1}\sum_{k=0}\frac{(r+1)^{(k)}}{k!} (\xi S)^{r+k-1}(-\tau)^k,
\label{eq:fmom_ser1}
\end{equation}
where $r^{(k)}$ denotes the rising factorial $r(r+1)\dots (r + k-1)$. For counter type II,  $\mu_\mathrm{[1]} = (1 + q)\xi S e^{-\tau\xi S}$, and its corresponding expression can be obtained from Eq.~(\ref{eq:fmom_ser1}) by replacing $r^{(k)}$ by $r^k$. That is, the expressions for types I and II are almost identical, differing only in the coefficients with the third and higher terms. These expressions form an alternating series what facilitate their convergence for a wide range of input parameters.

Let us substitute the expansion into the integral in Eq.~(\ref{eq:mandell}), thereby defining the factorial moments of the registered counts $\alpha_\mathrm{[r]}$. The integration of the variable $S$ provides the raw moments $\nu^\prime_j$ of the scintillation, and as a result, we have
\begin{equation}
\alpha_\mathrm{[r]} = G_r(1+q)^r\xi^r\sum_{k=0} \frac{r^{(k)}}{k!} (-\xi\tau)^k \nu^\prime_{r+k} + H_r(1+q)^{r-1}\xi^{r-1}\sum_{k=0}\frac{(r+1)^{(k)}}{k!}(-\xi\tau)^k\nu^\prime_{r+k-1}.
\label{eq:amom1}
\end{equation}
Accordingly, in the expression for counter type II, instead of using the rising factorial $r^{(k)}$, the power $r^{k}$ is substituted.

Evidently, with $\tau = 0$ and $q = 0$, these expressions reduce to the Mandel relation. From Eq.~(\ref{eq:amom1}), it follows that in the case of nonlinearity, the factorial moment of the registered counts is defined by the corresponding and subsequent raw moments of the scintillation with rapidly decreasing weights if $\xi\tau \ll 1$. After-pulses add a preceding raw moment to the sum.

We clarify Eq.~(\ref{eq:amom1}) for the example of $\alpha_\mathrm{[1]} \equiv \mathsf{E}[x]$ ($G_1 = 1, H_1 = 0 $) for the paralyzable dead time counter:
\begin{equation}
\alpha_\mathrm{[1]} = (1+q)\,\xi\sum_{k=0} \frac{(-\xi\tau)^k}{k!} \nu^\prime_{k+1} =\\= (1+q)\xi (1 - \tau\xi - s^2\tau\xi + \frac{1}{2}\nu^\prime_3\tau^2\xi^2 - \frac{1}{6}\nu^\prime_4\,\tau^3\xi^3 + \dots).
\label{eq:alpha1_dec1}
\end{equation}
Here, the moment $\nu^\prime_2$ is expressed via the scintillation index $s^2$, and $\nu^\prime_1 \equiv 1$. Setting all the values of $\nu^\prime_r = 1$, we obtain the standard formula for static nonlinearity: $\mathsf{E}[x] = \xi\,e^{-\tau\xi}$. We note that for strong scintillation ($s^2 \approx 1$), a significant additional bias appears in the form of dynamic nonlinearity, which amounts to the static nonlinearity.

\label{sec:normfmom}

The factorial moment $\alpha_\mathrm{[r]}$ is proportional to the value $\xi^r$, which is unknown in advance, and we have to somehow normalize the moments to avoid dependence on the brightness of the light source. For an ideal detector, normalization can be executed at once by using the value of $\xi^r$. In our case, the measured mean $\alpha_\mathrm{[1]}$ is more convenient. Let us denote the normalized moment by $\widetilde\alpha_\mathrm{[r]} = \alpha_\mathrm{[r]}/\alpha_\mathrm{[1]}^r$.

For a constant flux ($S \equiv 1$, the distribution $f(S; s^2)$ is degenerate, all the raw moments $\nu^\prime_{k} = 1$, and $\alpha_\mathrm{[r]} = \mu_\mathrm{[r]}$), all sums in the expression for $\widetilde\alpha_\mathrm{[r]}$ are reduced, and naturally, we obtain
\begin{equation}
\widetilde\alpha_\mathrm{[r]} = G_r + \frac{H_r}{\alpha_\mathrm{[1]}}.
\label{eq:amom_const}
\end{equation}
This equation indicates that detector nonlinearity has almost no effect on $\widetilde\alpha_\mathrm{[r]}$, and the effect of after-pulses should be very noticeable for low light intensities.

For stochastic radiation, $\widetilde\alpha_\mathrm{[r]}$ immediately provides an approximate estimate of the desired moment $\nu^\prime_{r}$ of the scintillation. Since the detector nonlinearity reduces the values of the numerator and denominator of $\widetilde\alpha_\mathrm{[r]}$, it is expected that the normalized moment is changed smaller. Indeed, in the linear approximation
\begin{equation}
\widetilde\alpha_\mathrm{[r]} = \nu^\prime_{r} - r(\nu^\prime_{r+1} - \nu^\prime_{r}\nu^\prime_{2})\,\tau\xi + \dots,
\end{equation}
the coefficient of $\tau\xi$ is certainly less than the corresponding coefficient in Eq.~(\ref{eq:amom1}).

\subsection{Alternative method to calculate low moments}

From Eqs.~(\ref{eq:probab1}--\ref{eq:mandell}) we can infer that for the calculation of the moment of a doubly stochastic process, it is not necessary to use the inner distribution $P(x; \xi S)$ if the corresponding moment is known. Further calculations are reduced to the weighted averaging of this moment with the outer distribution $f(S; s^2)$. In measurements of  scintillation (e.g., for probing of atmospheric optical turbulence), the most important values are the relative variance and covariance. Given the importance of these parameters, we  calculate the mean, variance and covariance of registered counts without using a specific form of the distribution $P(x; \xi S)$. This calculation is easy to perform for counter type II.

To obtain the mean of the registered flux, we use the well-established fact that
$\mathsf{E}[x; \xi S] = \xi S e^{-\tau \xi S}$. Therefore, we have
\begin{equation}
\alpha_1 = \mathsf{E}[x] = \int\limits_{0}^{\infty}\xi S e^{-\tau \xi S}\,f(S;s^2)\mathrm{\,d}S.
\label{eq:munl1}
\end{equation}
In addition, we consider the generating function of the central moments $\nu_k $ for the distribution $f(S; s^2)$ in the form
\begin{equation}
M_S(u) = \mathsf{E}[e^{u(S-1)}] = \int\limits_{0}^{\infty}e^{u(S-1)}\,f(S; s^2)\mathrm{\,d}S,
\label{eq:munl2}
\end{equation}
while assuming that it exists in the neighborhood of zero. It is well-known that this function and its derivatives are expressed through central moments by means of the following relationships:
\begin{equation}
M_S(u) = \sum_{k=0}^\infty \nu_k \frac{u^k}{k!}, \quad \frac{\partial{M_S(u)}}{\partial{u}} = \sum_{k=0}^\infty \nu_{k+1} \frac{u^k}{k!}, \dots
\label{eq:serz}
\end{equation}
Using $u=-\tau\xi$, it is easy to show from Eqs.~(\ref{eq:munl1}) and (\ref{eq:munl2}) that
\begin{equation}
\alpha_1 = e^{-\tau\xi}\Bigl(\xi\,M_S(-\tau\xi) - \frac{\partial{M_S(-\tau\xi)}}{\partial{\tau}} \Bigr) =\\= \xi\,e^{-\tau\xi}\Bigl(M_S(u) + \frac{\partial{M_S(u)}}{\partial{u}} \Bigr).
\label{eq:dertau}
\end{equation}
Substituting in this formula the series expansion of the MGF and its derivatives, i.e., Eq.~(\ref{eq:serz}), we obtain
\begin{equation}
\alpha_1 = \xi e^{-\tau\xi}\sum_{k=0}^\infty (\nu_k+\nu_{k+1}) \frac{(-\tau\xi)^k}{k!} =\\= \xi e^{-\tau\xi}\bigl(1 - \tau\xi\,s^2 + \frac{1}{2}\tau^2\xi^2(s^2 + \nu_3) - \dots \bigr).
\label{eq:alpha1}
\end{equation}
In the expanded formula, we replace $\nu_0 \equiv 1$, $\nu_1 \equiv 0$, and $\nu_2 \equiv s^2$. The resulting expression is slightly different in its form from Eq.~(\ref{eq:alpha1_dec1}) because the additional factor of $e^{-\tau\xi}$ is separated from the sum. The effect of after-pulses can be accounted by merely multiplying the equation with the factor $1 + q$.

To calculate the variance of the registered events, we use the well-known classical expression \cite{Feller1984-1e, Goldansky1959e} for variance in the case of a paralyzable counter: $\mathsf{Var}[k; \theta] = \theta e^{-\tau\theta}(1 - 2\tau\theta e^{-\tau\theta})$. Let us represent it in the form of $\mathsf{Var}[k] = \mathsf{E}[k] - 2\tau\mathsf{E}[k]^2$, which implies the equation
\begin{equation}
\mu_\mathrm{[2]} = \mu_\mathrm{[1]}^2(1-2\,\tau)
\label{eq:mu2cl}
\end{equation}
The factor of $(1-2\,\tau)$ is precisely the coefficient $G_2$, which has been introduced in Section \ref{sec:facmomGPD}. Using this equation, we can write
\begin{equation}
\alpha_\mathrm{[2]} = G_2\int\limits_{0}^{\infty} \xi^2 S^2 e^{-2\tau\xi S}\,f(S;s^2)\mathrm{\,d}S
\label{eq:f2int}
\end{equation}
This integral can be calculated in a manner similar to Eq.~(\ref{eq:munl1}) using the MGF $M_S(u)$. Omitting the intermediate transforms, we obtain
\begin{multline}
\alpha_\mathrm{[2]} = G_2\,\xi^2 e^{-2\tau\xi}\sum_{k=0}^\infty (\nu_k+2\nu_{k+1}+\nu_{k+2}) \frac{(-2\tau\xi)^k}{k!} =\\= G_2\,\xi^2 e^{-2\tau\xi}\bigl(1+s^2-2\,(2\,s^2+\nu_3)\,\tau\xi + 
\dots\bigr).
\label{eq:amom2cl}
\end{multline}
As in the case of Eq.~(\ref{eq:alpha1}), this formula does not exactly have the form of  Eq.~(\ref{eq:amom1}); however, in most cases, they coincide. Setting $\tau = 0$, we obtain $\alpha_\mathrm{[2]} =\xi^2 (1 + s^2)$, which after transition to variance, redices to known expression $\mathsf{Var}[x] = \xi + \xi^2 s^2$.

Based on the discussion in Sec~\ref{sec:afterpulse}, we account for the effect of after-pulses by multiplying the obtained expression by the factor of $(1 + q)^2$ and adding the term $2q\,\alpha_1$.

To calculate the covariance $\alpha_{12}$ of registered events, instead of using Eq.~(\ref{eq:fmom_calc0}), we write
\begin{equation}
\alpha_{12} = \mathsf{E}[x_1 x_2] =\\= \iint\limits_{0}^{\infty} \xi_1\xi_2 S_1 S_2e^{-\tau\xi_1 S_1 -\tau\xi_2 S_2}\,f(S_1,S_2))\mathrm{\,d}S_1 \mathrm{\,d}S_2,
\label{eq:cov_int}
\end{equation}
where the function $f(S_1, S_2))$ is a joint distribution of the correlated random variables $S_1$ and $S_2$. Owing to uncorrelated photon events, the joint distribution $P(x_1, x_2; \xi_1 S_1, \xi_2 S_2) = P (x_1; \xi_1 S_1) P(x_2; \xi_2 S_2)$, and the product of $\mathsf{E}[x_1; \xi_1 S_1]\mathsf{E}[x_2; \xi_2 S_2]$ appears under the integral. The variables $S_1$ and $S_2$ may represent scintillation values for different apertures (channels) or at different temporal points. In any case, the moment $\nu^{\prime}_{1,1} = \mathsf{E}[S_1 S_2]$ or $\nu_{1,1} = \nu^{\prime}_{1,1} - 1$ characterizes their correlation.

Similar to the approach assumed to calculate variance, we make use of the generating function of the joint central moments $M_S(u,v) = \mathsf{E}[e^{u(S_1-1)} e^{v(S_2-1)}]$, and we represent $\alpha_{12}$ in the form
\begin{equation}
\alpha_{12} = e^{-\tau_1\xi_1 -\tau_2\xi_2 }\Bigl(\xi_1\xi_2\,M_S(u,v) - \xi_2\frac{\partial{M_S(u,v)}}{\partial{\tau_1}} - \\ - \xi_1\frac{\partial{M_S(u,v)}}{\partial{\tau_2}} + \frac{\partial{^2M_S(u,v)}}{\partial{\tau_1}\partial{\tau_2}} \Bigr),
\end{equation}
where $u = -\tau_1 \xi_1$ and $v = -\tau_2 \xi_2 $. Function $M_S(u, v)$ and its derivatives can be expanded as in Eq.~(\ref{eq:serz}) in the series of joint moments $\nu_{k,l}$:
\begin{equation}
\alpha_{12} = \xi_1\xi_2\,e^{-\tau_1\xi_1 -\tau_2\xi_2 }\sum_{k=0}^\infty \sum_{l=0}^\infty \, (\nu_{k,l}+\nu_{k,l+1} +\\+ \nu_{k+1,l}+\nu_{k+1,l+1}) \frac{(-\tau_1\xi_1)^k}{k!}\frac{(-\tau_2\xi_2)^l}{l!}.
\label{eq:rho_fin}
\end{equation}
For clarity, we write the first few terms of this series:
\begin{equation}
\alpha_{12} = \xi_1\xi_2\,e^{-\tau_1\xi_1 -\tau_2\xi_2 } \bigl[1+\nu_{1,1} - (\nu_{0,2}+\nu_{1,1} + \nu_{1,2})\xi_2\tau_2 -\\- (\nu_{2,0}+\nu_{1,1} + \nu_{2,1})\xi_1\tau_1+ \dots\bigr].
\end{equation}
Subsequent to the subtraction of the products of the means, we note that the required moment $\nu_{1,1}$ is a key term in the series. The presence of after-pulses in channels 1 and 2 with respective probabilities of $q_1$ and $q_2$ is accounted for by the multiplicative factor $(1 + q_1)(1 + q_2)$.

\subsection{Convergence of expansions}

For practice, the inverse task, i.e., calculation of the scintillation moments on the base of measured moments $\widetilde\alpha_\mathrm{[r]}$, is more interesting. It greatly simplifies matters that only the scintillation moment of order $r$ enters into the right-hand side of the equations with weight nearly equal to unity. The following moments are included in the inverse calculation with small weights of the order of $\tau\xi$, $\tau^2\xi^2$, etc. In such a situation, we can use a simple iterative scheme starting from Eq.~(\ref{eq:mandell}) for  perfectly linear counters.

However, certain factors can complicate these calculations. For certain distributions of $f(S; s^2)$, raw moments quickly and indefinitely grow with their order, and therefore, decrease in $(\tau\xi)^k/k!$ does not guarantee convergence of the series. The sufficient conditions for the convergence of the alternating series given by Eq.~(\ref{eq:amom1}) are
\begin{equation}
\nu^{\prime}_{n+1} < \frac{n-r+1}{n\tau\xi}\nu^{\prime}_{n} \quad \mbox{and} \quad \nu^{\prime}_{n+1} < \frac{n-r+1}{r\tau\xi}\nu^{\prime}_{n},
\label{eq:cond1}
\end{equation}
for counters of types I and II, respectively, for all $n$ starting from some number. These inequalities indicate the following: 1) for counter type I, the restrictions are less stringent than for counter type II, 2) the convergence of the series depends on the nonlinearity $\xi\tau$ and a series converging at a small flux may diverge at a large flux, 3) convergence depends on order $r$, and it can only occur for the lowest moments.

For example, a log-normal distribution, which is often used as a representative distribution of stellar scintillation, has a moment $\nu^{\prime}_{n} = {\nu^{\prime}_{2}}^{n(n-1)/2}$. Therefore, Eq.~(\ref{eq:cond1}) leads to the condition ${\nu^{\prime}_2}^n < (n-r +1)/r\tau\xi$. Obviously, for any $r$ and arbitrarily small $\tau\xi$, the condition is never satisfied starting from some $n$, since $\nu^{\prime}_{2} > 1$ due to the normalization.

For the expansions given by Eqs.~(\ref{eq:alpha1}) and (\ref{eq:amom2cl}) as regards central moments, the conditions are similar, but include a combination of few consecutive moments. Thus, the series for $\alpha_\mathrm{[2]}$ converges if
\begin{equation}
\nu_{n+1} < \frac{n-r+1}{r\tau\xi}(\nu_{n}+2\,\nu_{n-1}+\nu_{n-2}).
\label{eq:cond2}
\end{equation}
Given that the central moments increase at a slower rate than raw ones, and that the right-hand side includes the sum of three consecutive moments, we expect that condition given by Eq.~(\ref{eq:cond2}) permits a wider range of initial data.

For practical purposes, a convergence in mathematical sense is not sufficient. In order to ensure that errors in measurements do not deteriorate a performance of the iterative scheme and its result, the actual convergence of the series should be provided to occur faster, for e.g., by considering only the linear and quadratic terms.

\subsection{Effect of background radiation}

Measured light may contain an additional constant component: background radiation $B > 0$. The background is described by a degenerate distribution in which all the central moments are equal zero except the mean $\beta = \mathsf{E}[B]$. We need to convert the previously obtained moments of the variable $S = (1-b)\,S^* + b$, where the relative fraction of the background $b=\beta/\xi$, to the moments of $S^*$. Obviously, $\mathsf{E}[S^*] = \mathsf{E}[S] = 1$. Using the properties of MGF, we obtain
\begin{equation}
M_{S-1}(t) = M_{S^*-1}((1-b)t),
\end{equation}
that leads to the equality $\nu_r = (1-b)^r\,\nu_r^*$ for the central moments. This implies that the results obtained in the previous step are transformed to the moments of the scintillation itself according to the relation
\begin{equation}
\nu_r^* = \frac{\xi^r}{(\xi - \beta)^r}\,\nu_r.
\label{eq:back}
\end{equation}

For raw moments, the equation can also be obtained from the MGF using the shift and scaling formula for a random variable:
\begin{equation}
M_{S^*}((1-b)t) = M_{S}(t)e^{-bt},
\end{equation}
the differentiation of which can provide the required relation. On the left-hand side, we obtain $(1-b)^r\,{\nu_r^\prime}^* $, and on the right, we obtain a polynomial of degree $r$ for $b$ and the moments $\nu_r^\prime$. Considering that $b \ll 1$, we get the approximate expression:
\begin{equation}
(1-b)^r\,{\nu_r^\prime}^* = \nu_r^\prime + rb\nu_{r-1}^\prime + \dots
\end{equation}
A similar relation is given in \cite{Jakeman1980TCP}, p.~87. From this equation, it follows that even the background radiation $b \sim 0.01$ can significantly affect higher moments when using the relation given by Eq.~(\ref{eq:back}).

\section{Experimental verification}
\label{sec:exper}

\subsection{Verification of basic distribution}

In the case of constant intensity, the moments $\alpha_\mathrm{[r]} = \mu_\mathrm {[r]}$, and thus, the distribution $P(x; \theta)$ can be verified experimentally. We carried out these measurements with the MASS instrument in the laboratory using an internal light source \cite{2007bMNRAS}. The experiment was conducted at a device temperature of about $29^\circ$C with photomultiplier tubes (PMTs) manufactured in 2003. As expected, these factors have led to a significant increase in the fraction of after-pulses. For each level of intensity over a wide range, the factorial moments of order $r = 1,2, \dots, 7$ were measured for an accumulation time of 60~s. A short exposure time of $\Delta t = 250\ \mu s$ was set in order to amplify desired effects.

\begin{figure}
\centering
\begin{tabular}{cc}
\psfig{figure=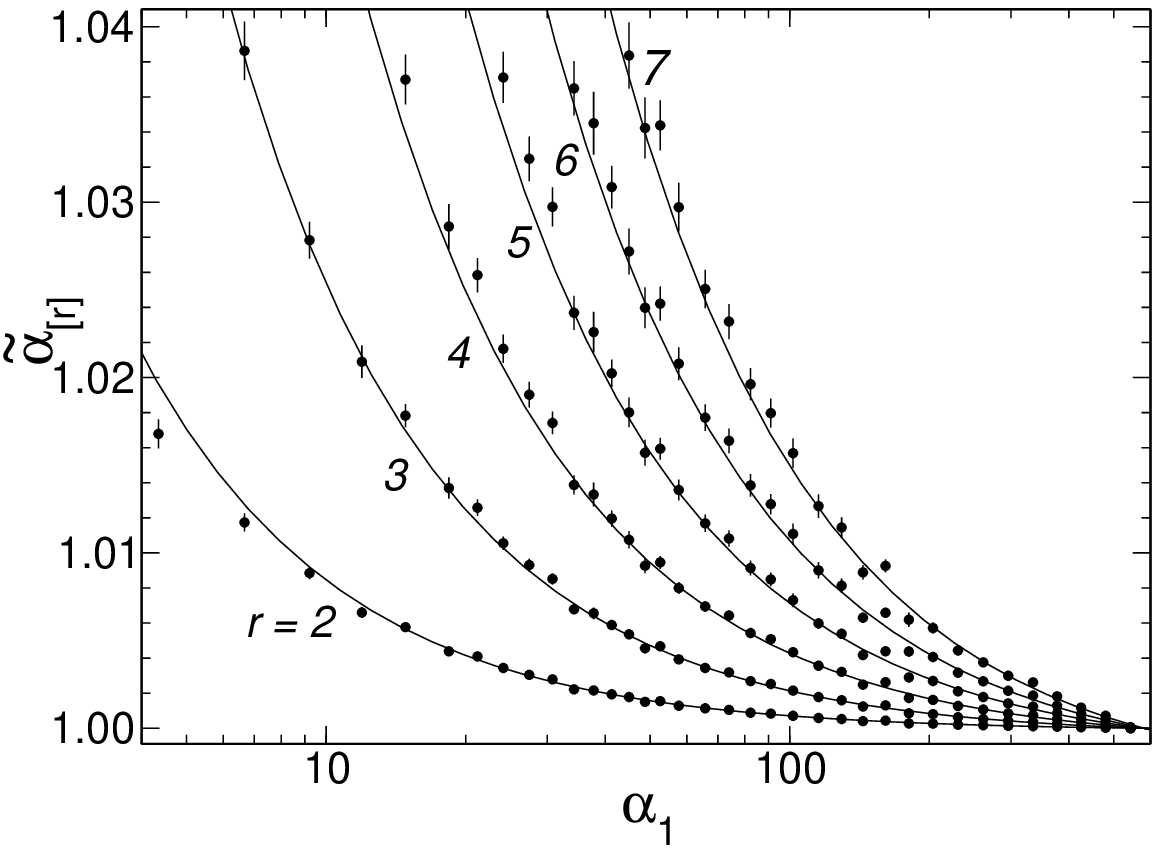,width=6.0cm,angle=-90} &
\psfig{figure=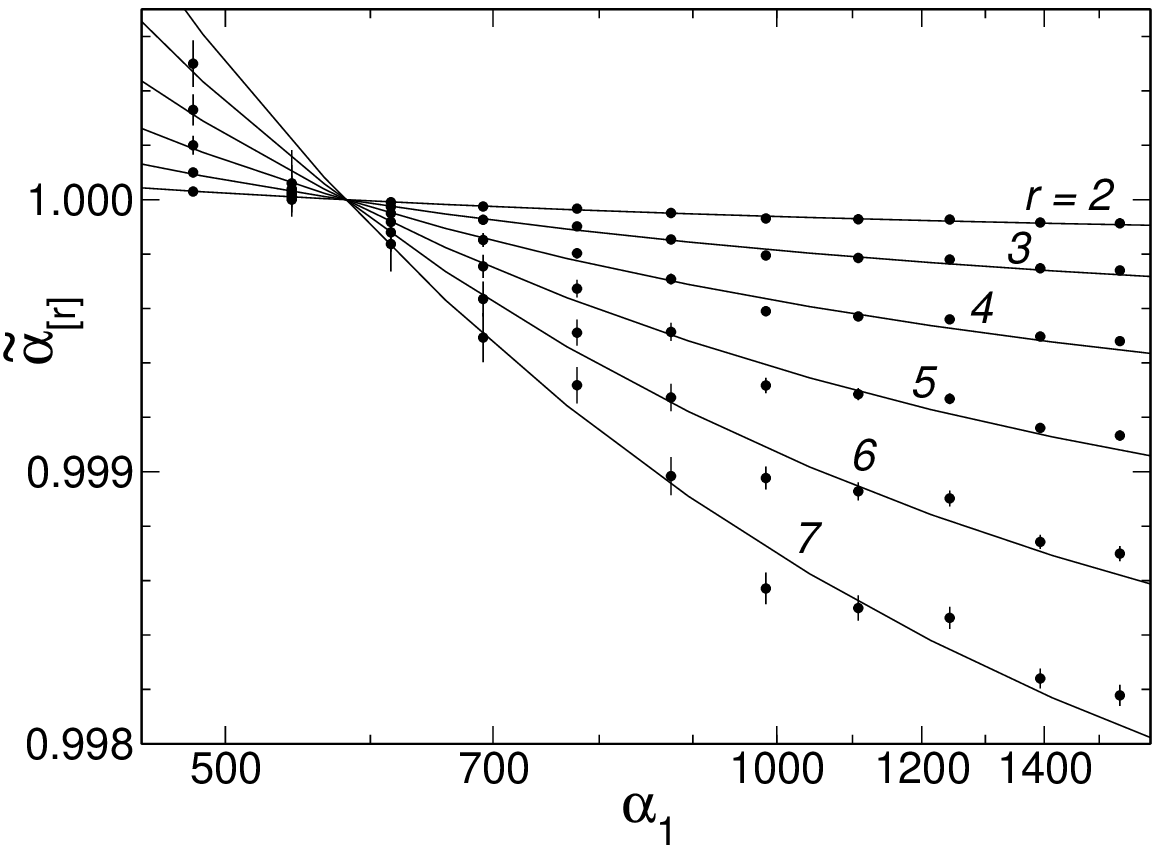,width=6.0cm,angle=-90} \\
\end{tabular}
\caption{Measured normalized moments $\tilde\alpha_\mathrm{[r]}$ of different orders $r = 2,\dots,7$ as a function of the mean $\alpha_\mathrm{[1]}$. Left: low intensity illumination. Right: large fluxes. The vertical segments indicate the estimate of the error of the measurement. The thin lines indicate the curves for each order $r$ as calculated using Eq.~(\ref{eq:amom_const}) with after-pulse probability $q = 0.043$ and $\tau = 7.4\cdot 10^{-5}$.
\label{fig:exp2}}
\end{figure}

The behavior of the measured normalized moments $\tilde\alpha_\mathrm{[r]}$ of various order $r = 2, \dots, 7$ depending on the measured mean $\alpha_\mathrm{[1]}$, is shown in the left panel of Fig.~\ref{fig:exp2} for low light fluxes. We note that the normalized moments increase with decreasing value of $\alpha_\mathrm{[1]}$ in agreement with Eq.~(\ref{eq:amom_const}). The right panel of the figure shows the dependence of $\widetilde\alpha_\mathrm{[r]}$ for large fluxes. All measured points lie below unity, as predicted by the theory.

Using measurements for $r = 3$, we determined the probability of after-pulses as $q = 0.043$. The value of the nonlinearity parameter $\tau = 7.4\cdot 10^{-5}$ was measured previously. The curves, constructed in accordance with the definition of the coefficients $G_r = 1 - r\,(r-1)\,\tau$ and $H_r = r\,(r-1)\,q/(1 + q)$ as per Eq.~(\ref{eq:amom_const}) are indicated  by thin solid lines. From Eq.~(\ref{eq:momGPD-after}), we note that the sum of the second and the third terms is equal to zero when $\tau\xi \approx q$ regardless of the order $r$. This fact is represented as the intersection of all the curves at one point with $\alpha_\mathrm{[1]} = 580$ and $\widetilde\alpha_\mathrm{[r]} = 1$.

The observed agreement between the theoretical description and the experimental data confirms that the distribution of registered events is close to the GPD and the accounting of after-pulses via Eq.~(\ref{eq:momGPD-after}) is valid. We note that the modified Poisson distribution \cite{Jakeman1980TCP, Vannucci1978OC} does not predict the observed behavior of $\widetilde\alpha_\mathrm{[r]}$ for large fluxes. The experimental results confirm our theoretical conclusions, and in particular, that the use of factorial moments at very low counts per exposure requires a guaranteed absence of the after-pulses.

The measurements shown in Fig.~\ref{fig:exp2} can be effectively used to determine the fraction of after-pulses $q$ with improved accuracy due to the large coefficient $r(r-1)$ for higher moments. It is noteworthy that in actual detectors, the value $q$ depends on several external factors that vary over time, and requires periodic inspection.

\subsection{Verification using data on stellar scintillation}

Numerous measurements of the statistical moments of stellar scintillation \cite{Jakeman1978,Parry1980,Dainty1982} show that, in fact, the moments are growing slower than the corresponding moments of the log-normal distribution. We also performed a series of measurements with the MASS instrument during monitoring of the optical turbulence in the atmosphere above Mount Shatdzhatmaz \cite{2010MNRAS} in the period January--April 2013. At this point, we recall that the device measures the fluctuations in light intensity with an exposure time of $\Delta t = 500\,\mu$s simultaneously in four entrance apertures A, B, C, and D, whose outer diameters are 2.0, 3.5, 6.3 and 9.0~cm, respectively. The nonlinearity parameters for all the channels are nearly identical, and $\tau = 36\cdot 10^{-6}$.

\begin{figure}
\centering
\psfig{figure=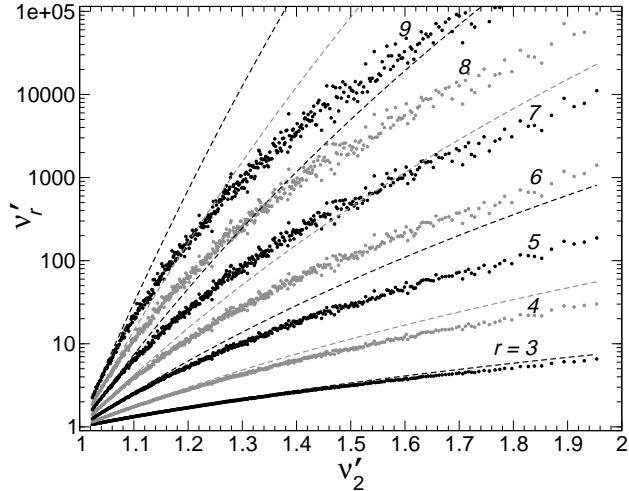,width=6.8cm,angle=-90}
\caption{Dependencies of measured moments $\nu_r^{\prime}$ ($r = 3, \dots, 9$) on the power of the scintillation in the form of $\nu^{\prime}_2$ for the 2~cm aperture. For comparison, the dashed lines show the dependencies $\nu^{\prime}_r(\nu^{\prime}_2)$ for the case of the log-normal distribution.
\label{fig:raw_mom}}
\end{figure}

Raw moments from the third up to the ninth, measured at the smallest aperture A, are shown in Fig.~\ref{fig:raw_mom} as a function of the second raw moment $\nu^{\prime}_2 = 1 + s^2$. The moments were calculated as $\nu^\prime_{r} \approx \widetilde\alpha_\mathrm{[r]}$, that is, without taking into account the nonlinearity of the detectors. Each point was calculated as an average of 20 estimates with an accumulation time of 1~s (40\,000 counts). This procedure  provides reasonably good statistical accuracy, and it allows the minimization of the impact of trends on the results. Although periods with very strong scintillations are rarely observed, our measurements cover a wide range of scintillation power. The figure illustrates the fact that in the case of strong scintillation, the actual instances of observed higher moments are considerably smaller than the moments of the log-normal distribution. For example, at $\nu^{\prime}_2 \approx 1.5$, the seventh moment $\nu^{\prime}_7$ is less than the
corresponding log-normal moment by an order of magnitude.

\begin{figure}
\centering
\psfig{figure=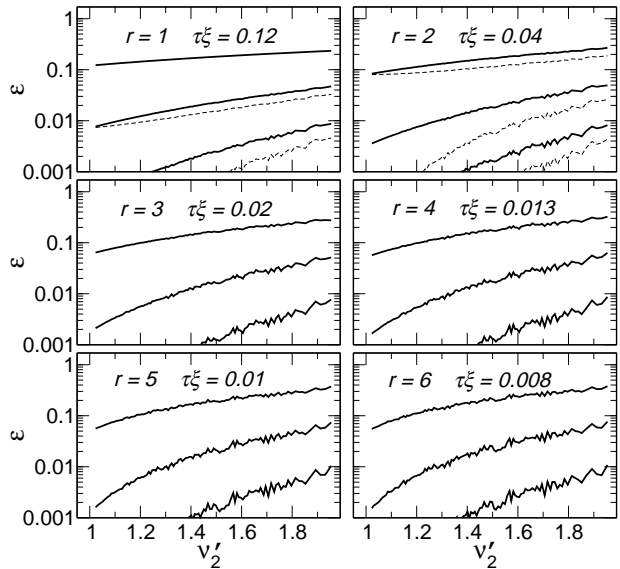,width=8.0cm,angle=-90}
\caption{Relative contribution $\varepsilon$ of consecutive terms of the series given by Eq.~(\ref{eq:amom1}) for the moments of order $r$, calculated using the scintillation data. The allowed nonlinearity parameters are specified in each panel.The thick curves depict the contributions of the linear, quadratic and cubic terms from top downward. The relative contributions for Eqs.~(\ref{eq:alpha1}) and (\ref{eq:amom2cl}) are shown by dashed curves. \label{fig:terms}}
\end{figure}

Using these data we tested the convergence of Eq.~(\ref{eq:amom1}) for moments of different orders. To perform this test, we attempted to determine for the nonlinearity $\tau\xi$ for which the relative error, introduced by a cubic term, amounts to $\approx 0.01$ in the worst case. The relative contributions of the first three terms, depending on the scintillation power, are shown in Fig.~\ref{fig:terms}. As might be expected, the allowable nonlinearity decreases rapidly with order and $\tau\xi$ becomes as low as 0.01 for $r=5$ which is usually considered to lie within the linearity range.

The contribution of the quadratic term is negligible (less than 0.01) under weak scintillation; however, it comes up to 0.05--0.08 under strong scintillation, and it should be taken into account. Linear term should always be considered, and under strong scintillation, it can be neglected only if the nonlinearity $\tau\xi$ is less than 0.001 even for the second moment. In our measurements, this condition corresponds to the mean $\xi < 40$ counts, what greatly impairs statistical accuracy.

With the same parameters, similar estimates are made for Eqs.~(\ref{eq:alpha1}) and (\ref{eq:amom2cl}). It is confirmed that these expansions converge more rapidly; for extremely strong scintillation, the contribution of the quadratic term is reduced by almost half, while the contribution of the cubic one becomes $\sim$0.005.

\subsection{Calculation of scintillation moments}

In this section, we first describe the series given by Eq.~(\ref{eq:amom1}) obtained under the assumption of GPD. The algorithm for the calculation of the required scintillation moment is sufficiently simple; we allocate the desired moment $\nu^\prime_{r}$ ($k = 0$) and the sum of corrections thereto $\Sigma_r$ ($k > 0$) in the right-hand side.

Because the measurements yield the normalized factorial moment $\widetilde\alpha_\mathrm{[r]}$, before substitution this moment must be multiplied by the factor $E^r$,  where $E$ represents the sum in Eq.~(\ref{eq:alpha1_dec1}), and it equals $\alpha_\mathrm{[1]}/\xi$ (see Sec.~\ref{sec:registered}). To calculate $\Sigma_r$, we use the measured $\widetilde\alpha_\mathrm{[r]}$ as estimates of the scintillation moments. By transposing the sum $\Sigma_r $ to the left-hand side, we obtain the corrected moment of the scintillation to the right. To assess the contribution of after-pulses, crude estimates of the moments are sufficient.

Under conditions of our measurements with MASS, $E$ is nearly unity, for e.g., for scintillation in the A and B apertures, this factor is $\sim$0.995, and for apertures C and D, it is about $0.99$, reaching $0.98$ for brightest stars. Therefore, the correction for detector nonlinearity is small (its relative value is only a few percent), and one iteration is sufficient for the calculation. However, relative corrections for higher moments ($r = 7, 8$) can be as large as $0.2$.

Despite the fact that the nonlinearity factor for aperture A is small ($\tau\xi \lesssim 0.004$), the corrections for A are comparable with the corrections for aperture D ($\tau\xi \lesssim 0.03$) because the scintillation in aperture D is approximately five times less intense.

In principle, for the comparison with theoretical behaviors, the dependencies shown in Fig.~\ref{fig:raw_mom} can be used without correction. The reason is that in the coordinates of this graph, the nonlinearity effect shifts data points along the measured dependencies. As result, the data points are not divided into subgroups, corresponding to weak and bright program stars, although the brightness of the stars differs more than 10 times.

\subsection{Calculation of scintillation power and covariance}

For the processing of the measurements of the mean, variance and correlation of the scintillation, Eqs.~(\ref{eq:alpha1}), (\ref{eq:amom2cl}) and (\ref{eq:rho_fin}) are better suited in practice. In these formulas, the effect of static nonlinearity is already separated out, and only the effect of dynamic nonlinearity requires decomposition. We recall that these formulas are valid for the counter with paralyzable dead time, which is a typical situation when using a photomultiplier as detector.

Let us denote the sum in Eq.~(\ref{eq:alpha1}) by $\Sigma_1$, and let us evaluate it using the estimates of the scintillation moments $\widetilde \alpha_\mathrm{[r]}$. The mean flux at the counter input can be calculated by using the formula from \cite{2008ARep}:
\begin{equation}
\xi e^{-\tau\xi} = \frac{\alpha_1}{(1+q)\Sigma_1}, \quad \xi\tau = -W\Bigl(-\frac{\tau\alpha_1}{(1+q)\Sigma_1}\Bigr),
\end{equation}
where $W(z)$ denotes the Lambert function, which is already known to us.

We select all the terms containing $\nu_2$ from the sum in Eq.~(\ref{eq:amom2cl}), and the residual of the sum is denoted as $\Sigma_2$ which is also of the order of unity. Dividing Eq.~(\ref{eq:amom2cl}) by $\alpha_1^2$, we obtain the expression for the normalized factorial moment:
\begin{equation}
\widetilde\alpha_\mathrm{[2]} = (1-2\,\tau)\frac{\nu_2\,(1-4\tau\xi+2\tau^2\xi^2) + \Sigma_2}{\Sigma_1^2} + \frac{2\,q}{\alpha_1}
\label{eq:normal}
\end{equation}
When using the factorial moments, the contribution of the Poisson noise is already accounted for, and the second term in the equation represents the difference between Poisson noise and  photon noise in the presence of after-pulses.

This equation is easily solved for $\nu_2$. Strictly speaking, $\nu_2$ is included in the term $\Sigma_1$ as well, and we should use an iterative process. To refine the higher moments we can use the general Eq.~(\ref{eq:amom1}). We note that $\nu_3$ is included in Eq.~(\ref{eq:amom2cl}) with weight $\sim$2$\tau\xi$ and its contribution to the actual measurements with MASS can range upto 3\% under strong scintillation conditions.

The double sum in Eq.~(\ref{eq:rho_fin}) for the covariance is of the order $1 + \nu_{1,1}$. As in the case of the variance, we split this sum into the terms with $\nu_{1,1}$ and the remainder $\Sigma_{12}$. Dividing this expression by the product of the averages $\alpha_{1(1)}\alpha_{1(2)}$, we obtain the following equation for the measured $\widetilde\alpha_\mathrm{12}$:
\begin{equation}
\widetilde\alpha_\mathrm{12}\Sigma_{1(1)}\Sigma_{1(2)} - 1 = \nu_{1,1}\,(1-\tau_1\xi_1-\tau_2\xi_2+ \tau_1\xi_1\tau_2\xi_2) + \Sigma_{12}.
\end{equation}
In the case of auto-covariance, it is possible to further simplify the formula since in this case, $\tau_1 = \tau_2$, $\xi_1 = \xi_2$. In addition, the relation $\nu_{k,l} = \nu_{l,k}$ holds for mixed moments.

\section{Conclusion}

In this study, we considered the impact of two well-known effects on the registration of  photodetector counts: dead time $\tau$ and after-pulses with probability $q$, which are inherent in different degrees in all counters of single-photon events. These effects distort the registered statistics in the measurements of the statistical characteristics of fluctuating radiation. Although our starting point is the measurement of stellar scintillation for remote sensing of the optical turbulence via the MASS measurement method, these relations are applicable to any similar problems, for example, even to those associated with fast and strong chaotic variability of physical and astronomical light sources.

Assuming that dead time leads to the distribution of counts (the number of detected photons per exposure) described by the generalized Poisson distribution, we derived general expressions for the factorial moments $\alpha_\mathrm{[r]}$ of the registered counts in the form of a series expansion with powers of the nonlinearity $\tau\xi$ combined with the raw moments $\nu^\prime_{k}$ ($k \ge r$) of light intensity.

Using well-known expressions for the nonlinearity of the counter with paralyzable dead time, we obtained the dependencies of the mean, variance, and covariance on the central moments of light scintillation. These expansions are more practical than the general ones, and they are not directly associated with a particular type of distribution of  photon events.

A detailed examination of these expansions proves that their convergence depends on the order of the factorial moment, the value of the nonlinearity $\tau\xi$, and the actual growth of scintillation moments. For processes with the log-normal distribution of  radiation, these series diverge at an arbitrarily small nonlinearity. The expansions indicate that under strong scintillation (when the relative variance of light fluctuations is of the order of unity), the dynamic nonlinearity owing to the curvature of the nonlinearity function is comparable to the static nonlinearity corresponding to its slope. In such a scenario, the usual treatment of the effect becomes insufficient.

The evaluation of the effect of after-pulses on the registered moments demonstrates that the contribution of after-pulses is usually underestimated. Even a small fraction ($\sim$1\%) of after-pulses leads to significant errors in the higher moments (of the order of tens of percent). At very weak fluxes, the usage of the factorial moments (Mandel relation) can lead to errors by several times.

We performed simple experiments to confirm the theoretically obtained relations. The results prove that the GPD adequately represents the distribution of photon counts in the case of nonlinearity caused by dead time during the registration process.

\end{document}